\documentclass[twocolumn,epsf,nofootinbib,superscriptaddress]{revtex4}
\usepackage{graphicx}% Include figure files
\usepackage{amsmath,amssymb,latexsym}
\usepackage{bm}% bold math
\usepackage{float} %necessary when used with multicols

\def\lbldef#1#2{\expandafter\gdef\csname #1\endcsname {#2}}

\def\href#1#2{#2}

\newcommand{\bwide}{\begin{widetext}}
\newcommand{\ewide}{\end{widetext}}
\newcommand{\beq}[1]{\begin{equation} \label{(#1)}}
\newcommand{\eeq}{\end{equation}}
\newcommand{\ba}[1]{\begin{eqnarray} \label{(#1)}}
\newcommand{\ea}{\end{eqnarray}}

\begin{document}

\renewcommand{\thefootnote}{\fnsymbol{footnote}}
\setcounter{footnote}{0}

\title{A Galactic Halo Origin of the Neutrinos Detected by IceCube}

\author{Andrew~M.~Taylor}
\affiliation{Dublin Institute for Advanced Studies,
31 Fitzwilliam Place, Dublin 2, IRELAND}

\author{Stefano~Gabici}
\affiliation{APC, AstroParticule et Cosmologie, Universite Paris Diderot, 
CNRS, CEA, Observatoire de Paris, Sorbonne Paris Cite, FRANCE}

\author{Felix~Aharonian}
\affiliation{Dublin Institute for Advanced Studies,
31 Fitzwilliam Place, Dublin 2, IRELAND}
\affiliation{Max-Planck-Institut f\"ur Kernphysik, 
             Postfach 103980, D-69029 Heidelberg, GERMANY}

%\date{}
\begin{abstract}
%\begin{center}
Recent IceCube results %\cite{Aartsen:2013bka,Aartsen:2013jdh} 
suggest that the first detection of very high energy astrophysical neutrinos have been 
accomplished. We consider these results at face value in a Galactic origin
context. Emission scenarios from both the Fermi bubble and broader halo region are 
considered. We motivate that such an intensity of diffuse neutrino emission could be 
Galactic in origin if it is produced from an outflow into the halo region. This 
scenario requires cosmic ray transport within the outflow environment to be different 
to that inferred locally within the disk and that activity in the central part of
the Galaxy accelerates cosmic rays to trans-''knee'' energies before they escape into 
an outflow. The presence of a large reservoir of gas in a very extended halo around the 
Galaxy, recently inferred from X--ray observations, implies that relatively modest 
acceleration power of $10^{39}$~erg~s$^{-1}$ in PeV energy cosmic rays may be sufficient 
to explain the observed neutrino flux. Such a luminosity is compatible with that required 
to explain the observed intensity of CR around the ``knee''.
%\end{center}
\end{abstract}

\maketitle

\section{Introduction}
\label{Introduction}

% 28 events observed with a background expectation of 10
The IceCube collaboration has recently reported the detection of 28 neutrinos with energies 
in excess of $\approx 30$~TeV, on an expected background of $10.6^{+5.0}_{-3.6}$ events. 
A purely atmospheric origin for the detected events has thus been rejected at the $4~\sigma$ 
level \cite{Aartsen:2013bka,Aartsen:2013jdh}. Data have been accumulated from some 662~days 
of observation of the full sky. Furthermore, although limited in statistics,
the neutrino distribution indicates a very extended if not isotropic distribution of 
arrival directions of these neutrinos \cite{Aartsen:2013jdh,Ahlers:2013xia}.%pg 4

%an excess of $\sim$18 neutrino events above an expected background
%of 10 events accumulated from some 662~days of observation of the 
%full sky 

Such an excess of events corresponds to a diffuse energy flux of neutrinos in the energy 
interval 0.1-1~PeV, for all three flavours, at the level:
\begin{eqnarray}
\label{eq:IceCube}
E_{\nu}^{2}\frac{dN}{dE_{\nu}}\approx 30{\rm ~eV~cm}^{-2}{\rm ~s}^{-1}{\rm ~sr}^{-1},
\end{eqnarray}
with a spectral slope which is estimated to be close to flat (i.e. $\alpha\approx 2$ for 
$dN/dE_{\nu}\propto E_{\nu}^{-\alpha}$) in this representation. 

The origin of the neutrinos detected by IceCube presently remains unknown. Both
Galactic \cite{Ahlers:2013xia,Neronov:2013lza,Lunardini:2013gva} and extragalactic 
\cite{rolandneutrinos,muraseextragalactic} scenarios of their production have been 
proposed, with a tendency to disfavor Galactic models other than those involving 
a connection with the Fermi bubble structures. These structures, whose existence were 
only recently disclosed both in $\gamma$-rays and radio \cite{Su:2010qj,Carretti:2013sc}, 
extend well outside the Galactic plane region and may well house a significant 
population of cosmic ray (CR) particles.

%\section{Galactic Plane Emission}

Generally, two classes of scenarios can be envisaged in an attempt to explain 
the apparently isotropic neutrino flux detected by IceCube. The observed flux level 
(Eq.~\ref{eq:IceCube}) might either result from the superposition of discrete 
sources or be truly diffuse on some scale. Indeed, insight into the problems facing
the origin of this emission may be obtained through the consideration specifically 
of one of these scenarios.

Assuming that some fraction of the neutrino flux recently observed 
is not actually diffuse on the largest scales, originating instead from the 
Galactic plane region, an indication of the expected neutrino detection rate can be 
derived from the $\gamma$-ray emission flux from this region,
some $\Omega_{\rm d}\sim 0.1$~sr in size. The level of very high energy 
$\gamma$-rays allowed from the Galactic plane, as was 
considered in \cite{Gabici:2008gw}, the MILAGRO observations \cite{Abdo:2008if} 
which partially covered this region provide a basis for determining its multi-TeV 
gamma-ray brightness. Using these observations to determine the corresponding 
neutrino flux brightness from the Galactic plane, the expected detection rate 
of neutrinos from the region can be determined.
Specifically, the MILAGRO observations from this region, whose median photon energy 
was estimated to be 15~TeV, motivate a photon energy flux at the level,
\begin{eqnarray}
E_{\gamma}F_{\gamma}^{\rm d}=\Omega_{\rm d}E_{\gamma}^{2}\frac{dN}{dE_{\gamma}}\approx 70{\rm ~eV~cm}^{-2}{\rm ~s}^{-1}.
\end{eqnarray}
%Assuming the origin of this emission comes from the Galactic plane region
%with an average distance $d_{s}\approx 8{\rm ~kpc}$, 
%%subtending a solid angle $\Delta\Omega\approx 0.1$~srad.,
%\begin{eqnarray}
%L_{\nu}=4\pi d_{s}^{2}E_{\nu}F_{\nu}=4\times 10^{35}~{\rm ~erg ~s}^{-1}.
%\end{eqnarray}
%Adopting a scale height of 1~kpc for the Galactic plane emission, and 
%a proton number density of the molecular material in the disk of 1~cm$^{-3}$,
%$t_{\rm esc}/t_{pp}=(3\times 10^{5})/(4\times 10^{7})=0.008$, and a corresponding proton 
%luminosity of $L_{p}\approx 2\times 10^{38}$~erg~s$^{-1}$ is required.
For the highly optimistic scenario in which the spectrum of parent protons continues 
with an $E^{-2}$ spectral shape up to a cutoff energy of 30~PeV, the corresponding
neutrino detection rate expected from the Galactic plane region is obtained by 
convolving the parent CR flux with the IceCube effective area.
For the effective area, we adopt a monotonic function of the form
\begin{eqnarray}
%A_{\rm eff}\approx 10^{0.7} \left(\frac{E}{E_{0}}\right)^{3.4}\prod_{i=1}^{3}\left(\frac{E_%{i}}{E_{i}+E}\right){\rm ~m}^{2}
A^{\nu}_{\rm eff}\approx A_{0} \left(\frac{E}{{\rm TeV}}\right)^{\gamma}e^{-(E_{b}/E)}{\rm ~m}^{2}.
\end{eqnarray}
%with $E_{0}=1$~GeV, $E_{1}=240$~GeV, $E_{2}=310$~GeV, and $E_{3}=37$~TeV.
with $A_{0}=$1, 0.9, 0.4, $\gamma=$0.4, 0.4, 0.5, and 
$E_{b}=$117~TeV, 155~TeV, 170~TeV for 
$\nu_{e}$, $\nu_{\mu}$, and $\nu_{\tau}$, respectively. This parameterisation
is found to fit well, within an accuracy of $\sim 20\%$, 
the published all--sky effective area for the three
different neutrino species shown in fig.~7 of \cite{Aartsen:2013jdh}.
Thus, overall, a total of $\sim$1 event per year is predicted from the Galactic plane
region, with $>$30~TeV energy neutrinos dominating the contribution to this rate.

%\begin{figure}
%\includegraphics*[width=0.5\linewidth, angle=-90]{secondary_g_GP.ps}
%\includegraphics*[width=0.5\linewidth, angle=-90]{secondary_nu_rates_IC_GP.ps}
%\caption{TOP: The gamma-ray energy flux for a scenario in which the MILAGRO
%observations of the Cygnus region are explained by a CR population with shape 
%$dN/dE\propto E^{-2}e^{-E/E_{\rm max}}$, with $E_{\rm max}=$10 and 30~TeV respectively. 
%BOTTOM: The corresponding Galactic plane neutrino detection rate by IceCube, 
%following the assumption that the multi-TeV Cygnus region/Galactic plane respective
%brightness levels follow the same relation as that of Fermi GeV fluxes from these 
%regions.}
%\label{Galactic_Plane_Emission}
%\end{figure}

This result clearly demonstrates the inability for this bright diffuse source 
to account for the level of flux apparently observed in neutrinos
at multi-TeV energies (see also \cite{Guo:2013rya}).
It should be noted that the numbers obtained above are
a factor of $\sim$2 smaller than those obtained in \cite{Razzaque:2013uoa} for the 
Galactic center region, for which an $E^{-2.3}$ power-law scaling from the 100~GeV fluxes, 
at the level 3500~eV~cm$^{-2}$~s$^{-1}$~sr$^{-1}$ observed by Fermi \cite{FermiLAT:2012aa},
were adopted. 
%Interestingly, the MILAGRO result scaled for the Galactic center region does approximately agree with that obtained following such power-law scaling from the Fermi flux levels.

%%%%%%%%%%%%%%% BEGIN

%These two scenarios are briefly discussed below, 
%where the additional assumption is made that the neutrinos originate within 
%the Galaxy. %and are produced through inelastic proton--proton interactions.
%IS P-GAMMA OF ANY RELEVANCE?

More generally, as shown in \cite{Gabici:2008gw,vissani2011}, the 
following simple but  nevertheless robust rule--of--the--thumb can be used: 
a neutrino flux at the level corresponding in $\gamma$-rays to 1~Crab 
(i.e. $F_{\nu}(> 1~{\rm TeV}) > 10^{-11} \nu{\rm ~cm}^{-2}{\rm ~s}^{-1}$) 
would yield a detection rate of about 1 neutrino per flavor per year 
in a detector whose size is one cubic kilometer. For typical spectra of astrophysical 
sources, the count rate is dominated by $\approx 10$~TeV neutrinos, and 
decreases at larger energies. This implies that a quite large number of 
discrete neutrino sources with fluxes at the Crab level (or, equivalently, 
an unreasonably large number of significantly weaker sources) are required 
to explain IceCube data. If neutrinos are produced through inelastic proton--proton 
interactions, a $\gamma$-ray flux of the same order of magnitude is also 
expected from such sources. Thus, the scarce number of very high energy 
$\gamma$-ray sources detected by current instruments at the Crab flux level 
seems to rule out this possibility. In the same context, a detailed 
investigation has been recently performed to assess the possible contribution 
to the neutrino flux from unidentified TeV sources in our Galaxy 
\cite{meszaros} and the results from this study are in line with the simple 
considerations made above.

%%%%%%%%%%%%%%%% END

%%%%%%%%%%%%%%%% BEGIN

A possible way out is to invoke the existence of a population of heavily 
absorbed $\gamma$-ray sources in the Galaxy. The most effective 
absorption mechanism for $\gamma$-rays in astrophysical environments 
is pair--production in a soft photon 
field. The presence of an intense radiation field would dramatically 
suppress the $\gamma$-ray flux from a source, while leaving the neutrino 
flux unaffected. 
%Reduction in the flux through $\gamma - \gamma$ absorption before escape from 
%the source, may therefore alleviate the above mentioned restrictions placed on 
%the candidate sources. 
Naturally, such a scenario may only be played out in 
compact objects like binary systems \cite{teresafelix} or hypernovae, which
have also been considered as candidate neutrino sources \cite{meszaros}.

%Though it might require quite extreme conditions to be 
%satisfied at source, such a scenario cannot be ruled out (see e.g. 
%\cite{teresafelix}), and is therefore worthy of further consideration. 
%This scenario will briefly be addressed in section \ref{Galactic_Plane}.

%%%%%%%%%%%%%%% END

%Within the context of the neutrino emission originating from the Galactic
%plane region, it is also worth addressing the possibility that the flux 
%originates directly from CR interactions within a Galactic source population. 

%Although the $\gamma$-ray fluxes detected in the Galactic plane region place
%strong constraints on possible neutrino sources, one can avoid a conflict with
%these observations provided that the flux is strongly absorbed before arriving. 

Finally, regardless of whether the source
of the emission is generated through CR interactions within their actual sources or via 
their interactions with atomic and molecular gas material in the disk, the arrival directions of 
the neutrinos produced are expected to originate from the Galactic plane region. Though 
present observations are consistent with a broader than disk distribution for the arriving
neutrinos, only through an improvement in statistics can a deeper probe of the underlying 
flux distribution be made.

Do the above considerations exclude a Galactic origin of the reported flux of PeV
neutrinos? No. In this paper we argue that a Galactic origin of these neutrinos remains
a viable option if one assumes that they are produced in the Galactic halo. This model
assumes that the neutrinos result from PeV CR interactions, after their escape from the 
Galactic disk, with the diffuse ambient gas of non-negligible density present, giving
rise to a quasi-isotropic neutrino flux at the level detected by IceCube.
In section~\ref{plane_halo_ratio}, a comparison is
made of the relative neutrino emission rates from the Galactic plane and halo regions, 
under the constant CR intensity assumption. 
In section~\ref{neutrino_production} we address the timescales involved for both CR escape
from the Galaxy and their energy loss times, which collectively dictate the
efficiency with which power is converted from CR to neutrinos.
In section~\ref{results}, non-uniform CR intensity scenarios in which CR within Galactic 
outflows power diffuse neutrino flux emission are put forward. In 
section~\ref{HAWC_LHAASO}, we consider the prospects for testing such scenarios using
the associated electromagnetic emission expected to accompany that output in neutrinos.
We summarise out conclusions in section~\ref{conclusion}.

%ALSO DISCUSSION OF TAVANI CYG-X3 FLUX CALCULATION SHOULD BE INCLUDED- FIND REF?

%Furthermore, it should be borne in mind that there is little motivation for
%an extension of the MILAGRO level flux with an $E^{-2}$ spectral shape up
%to tens of PeV in energy. However, since the scenario may be testable in
%the near future by upcoming experiments, it seems worthwhile to consider
%such a possibility.

%\subsection{CR Diffusion Emission in a Large Galactic Halo}

\section{Diffuse Plane/Halo Emission Ratio}
\label{plane_halo_ratio}

We here discuss the case in which the observed neutrino flux is truly diffuse 
and originates from the interactions between CR and ambient gas 
in the Galaxy.
The intensity $dN/dE_{\rm CR}$ of the CR responsible for the production of the 
high energy neutrinos is assumed to be constant throughout the whole Galaxy 
(disk plus halo). Such a setup can be considered as a zeroth order approximation 
for the possible configuration on which further considerations will be based. Under such 
circumstances, the expected number of neutrinos $N_{\nu}$ detected by a given 
telescope in a time $\Delta t$, from a region subtending a solid angle 
$\Delta \Omega$, and within an energy interval $\Delta E_{\nu}$ can be written 
as:
\begin{equation}
\label{eq:Nnu}
N_{\nu} ~\propto~ \sigma_{pp} ~ \frac{dN}{dE_{\rm CR}} ~ N_H ~ \Delta \Omega ~ \Delta E_{\nu} ~ \Delta t
\end{equation}
where $\sigma_{pp}$ is the relevant cross section and $N_H = n_{p} L$ is 
the gas column density along the line of sight. Here, $L$ is the length 
of the line of sight characterized by a typical interstellar hydrogen density $n_p$ 
and, as an order of magnitude estimate, in the following discussion it 
will be considered equal to the size of the considered system. 

Typical values for the gas density and the size of the disk are 
$n_p^d = 1 ~ n_{p,0}^d$~cm$^{-3}$ and $L^d = 10 ~ L^d_1$~kpc, which give a column 
density of $N_H^d = 3 \times 10^{22} n_{p,0}^d L^d_1$~cm$^{-2}$, while for the 
halo the following reference values are adopted, 
$n_p^h = 10^{-3} ~ n_{p,-3}^h$~cm$^{-3}$ and $L^h = 10 ~ L^h_1$~kpc which gives a column 
density equal to $N_H^h = 3 \times 10^{19} n_{p,-3}^h L^h_1$~cm$^{-2}$. 
Following \cite{Gabici:2008gw} we consider an extension for the disk in 
Galactic longitude and latitude of $-40^{\circ} < l < 40^{\circ}$ and 
$-2^{\circ} < b < 2^{\circ}$, respectively, whose corresponding solid angle is 
then $\Delta \Omega^d \approx 0.1$~sr, while for the Galactic halo we adopt
an optimistic value of $\Delta \Omega^h = 2 \pi$~sr.
By using Eq.~\ref{eq:Nnu} it is now possible to compute the ratio between 
the number of neutrinos detected from the halo and those detected from the 
disk of the Galaxy. This reads:
\begin{equation}
\frac{N^h_{\nu}}{N^d_{\nu}} ~=~ \left( \frac{n_p^h}{n_p^d} \right) \left( \frac{L^h}{L^d} \right) \left( \frac{\Delta \Omega^h}{\Delta \Omega^d} \right)
\end{equation}
$$
\approx 0.05 \left( \frac{n_{p,-3}^h}{n_{p,0}^d} \right) \left( \frac{L_{1}^h}{L_{1}^d} \right) \left( \frac{\Delta \Omega^d}{0.1} \right)^{-1} ~ .
$$
It follows that under the constant CR intensity assumption, the diffuse neutrino 
flux is dominated by the neutrinos coming from the Galactic disk, unless the 
Galaxy has a very extended halo 
(i.e. $L_1^h \ll L_1^d$). The recent claim for the detection of a huge 
reservoir of ionized gas in a $\approx 100$~kpc region around the Milky 
Way \cite{Gupta:2012rh} might give support to the latter scenario. Given that 
the neutrino emission from the Galactic disk might hardly explain the 
IceCube data (see \cite{Gabici:2008gw} and the discussion in 
Sec.~\ref{Introduction}), it seems more plausible to consider an 
extended Galactic halo as the site of production of the observed neutrinos.

To summarize, the simple considerations made above under the constant CR
flux intensity approximation motivate that the Galactic halo can potentially
be an important source of Galactic neutrinos.
For this  case, some natural candidate production sites are the Galactic halo 
itself (if sufficiently extended) and the Fermi Bubbles, which are giant structures, 
subtending $\Omega_{\rm FB}\sim 0.8$~sr, detected in GeV gamma rays that extend for 
several tens of kiloparsecs away from the Galactic disk.

In the following sections, we consider further the possibility that the
arriving neutrino flux consists of large scale diffuse emission. 
For this scenario, we determine the required CR luminosity levels needed to 
support a flux at the level recently measured.

\section{Source Luminosity Requirements}
\label{neutrino_production}

The secondary neutrino and parent CR luminosities are related by
\begin{eqnarray}
L_{\nu}=f L_{\rm CR},
\end{eqnarray}
where $L_{\nu}$ and $L_{\rm CR}$ are the neutrino and CR luminosities
and $f$ the efficiency of energy transfer between the CR and neutrino
populations. Such an efficiency relates to the fractional energy passed
into the neutrino population through inelastic proton--proton interactions, 
$K_{\nu}$, the energy loss time, $t_{pp}$, and the escape time $t_{\rm esc}$,
\begin{eqnarray}
f=K_{\nu}(1-e^{-\frac{t_{\rm esc}}{t_{pp}}}).
\end{eqnarray}

\begin{eqnarray}
t_{pp}=4\times 10^{9}\left(\frac{10^{-2}{\rm ~cm}^{-3}}{n_{p}}\right){\rm ~yrs},\label{t_pp}
\end{eqnarray}
and
\begin{eqnarray}
\label{eq:tesc}
t_{\rm esc}=3\times 10^{9}\left(\frac{R}{100{\rm ~kpc}}\right)^{2}\left(\frac{10^{30}
{\rm ~cm}^{2}{\rm ~s}^{-1}}{D}\right){\rm ~yrs}\label{t_esc}
\end{eqnarray}
where $D$ is the energy dependent CR diffusion coefficient. Eq.~\ref{eq:tesc} 
implicitly assumes 
that the transport of CRs proceeds in the diffusive regime. If, on the other hand, 
an advective flow of velocity $u_{\rm adv}$ dominates over diffusion, a more appropriate 
expression is 
\begin{eqnarray}
t_{\rm esc} = 3\times 10^{9}\left(\frac{R}{100~{\rm kpc}}\right)\left(\frac{30~{\rm km~s}^{-1}}{u_{\rm adv}}\right){\rm yrs},
\end{eqnarray}
in which case the escape time is expected to be independent of the particle energy.

Thus, for a given size region $R$, the underlying CR luminosity
required to support the inferred neutrino luminosity may be determined.
%So far in the discussion, no scenario for the actual injection of multi-PeV 
%cosmic rays into the Galactic halo region has been addressed. 
In the following sections we will specifically consider whether different
scenarios for a Galactic outflow emission origin are able to explain the 
observed flux.

Before proceeding, however, it is helpful to highlight the following general point
about CR interactions with Galactic material.
Following Eq.~\ref{t_pp}, for interaction times on the size of the Hubble scale, 
the number density of target material required is $n_{p}\approx 10^{-3}{\rm ~cm}^{-3}$.
Similarly, setting the escape time from the Galactic halo, given in Eq.~\ref{t_esc},
to this scale requires a diffusion coefficient within the halo of 
$D\approx 10^{29}{\rm ~cm}^{2}{\rm s}^{-1}$. Thus provided that the diffuse halo region
is sufficiently turbulent (ie. $\delta B/B\sim 1$) so as to support a diffusion 
coefficient at this level for CR up to a given energy, the Galaxy can be expected to 
operate as a calorimeter for these particles. Note that the diffusion coefficient of 
$\approx 10^{29}{\rm ~cm}^{2}{\rm s}^{-1}$ quoted above corresponds to the Bohm diffusion 
coefficient of $\approx 10$~PeV particles in a highly turbulent magnetic field 
of few microGauss with a coherence length greater than a few pc. 
Such a set-up also requires that advective escape is sufficiently 
slow so as not to provide shorter escape times.

We next consider specific Galactic origin scenarios in order to determine the
expected level of neutrino emission from these regions given what we have
already learnt from them through investigations of them in $\gamma$-rays.
For the description of the $\gamma$-ray yields following $pp$ interactions and 
their energy distribution, the \cite{Kelner:2006tc} parameterisation is adopted.

\section{Galactic Outflow Emission}
\label{results}

If diffuse on larger angular scales than the Galactic plane, the detection of 
surprisingly bright neutrino emission at PeV energies can have important 
implications with regards their Galactic ejection and escape at multi-PeV 
energies. Indeed, in the following section we consider a departure from the
constant intensity assumption for CR throughout the Galaxy, with Galactic
activity from/near the central region, either the central blackhole itself 
\cite{Cheng:2011tx} or that from a nearby central starburst region 
\cite{Crocker:2010qn}, powering fast CR acceleration and advection 
into the Galactic halo region. It is worth highlighting that CR
which enter an advective flow are not expected to return to the disk region,
and therefore, with regards spallation constraints on their propagation time, 
can be considered to have effectively escaped \cite{Ptuskin:1997}.

%Such an outflow scenario may motivate the presence of a 
%fresh young population of multi-PeV CR ejected out of the Galactic plane 
%following recent AGN activity. Interestingly, this scenario can also naturally 
%connect the flux level detected from the Fermi bubbles with the diffuse neutrinos 
%observed. On the other hand, with the Fermi bubbles only a factor of a few 
%brighter than the diffuse background level, the actual emission region may be 
%much larger than the observed bubbles. In this case, the presence of a much
%older multi-PeV CR population in the halo region may also be the origin of the 
%detected neutrinos.

{\bf Fermi Bubbles:} recent ejection from the Galactic center in the last 
few Myr into the Fermi bubble region may have deposited a fresh population 
of CR which have not had time to diffusively escape from the region. Such 
a scenario apparently fits in with recent dynamic modelling of the Fermi 
bubble structures \cite{Barkov:2013gda}.

With constraints on the possible multi-TeV $\gamma$-ray flux from
this region being placed by extrapolations from Fermi satellite measurements,
an optimistic estimate of the number of neutrinos expected from the Fermi bubble
regions may be obtained.

\begin{figure}
\includegraphics*[width=0.5\linewidth, angle=-90]{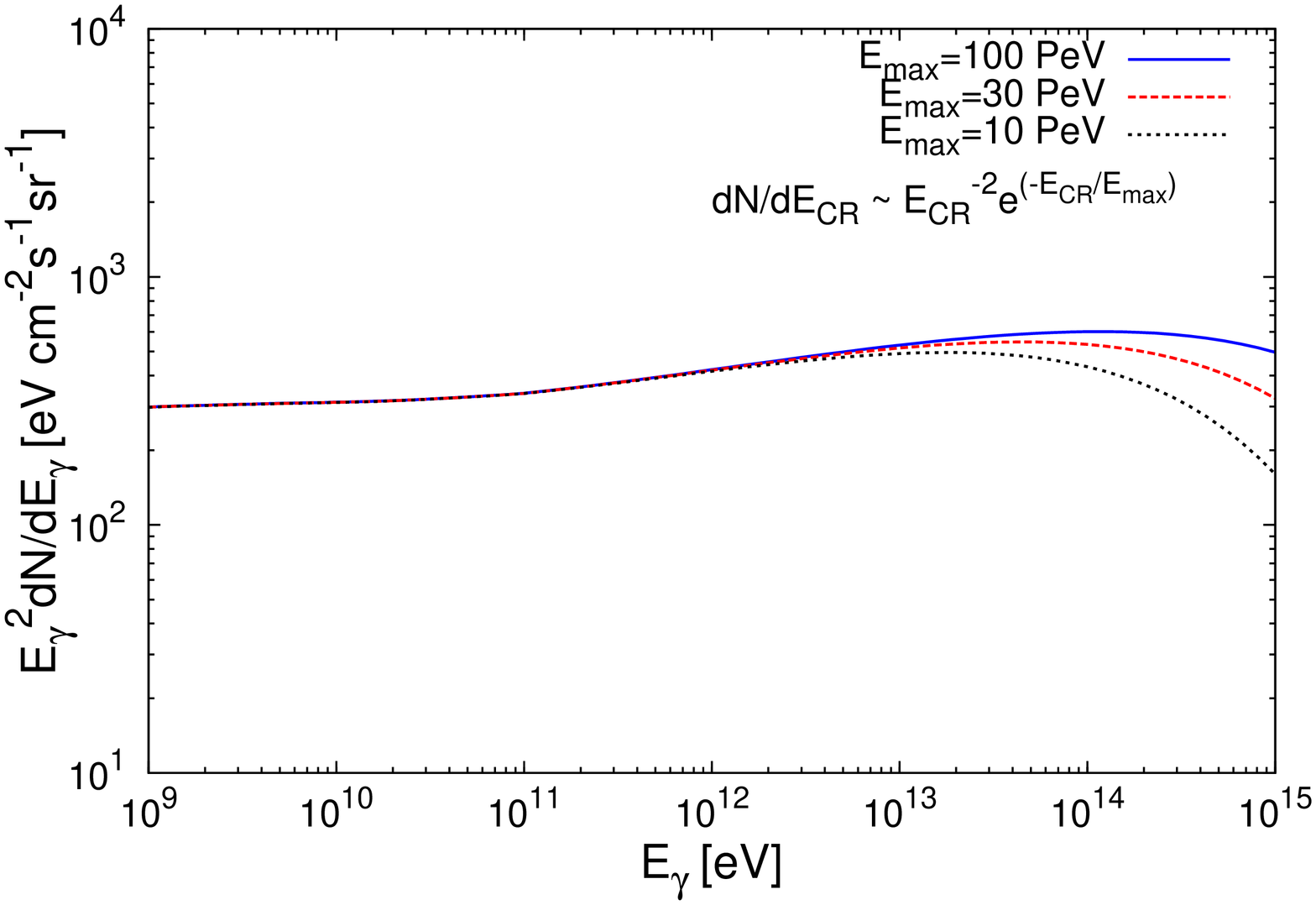}
\includegraphics*[width=0.5\linewidth, angle=-90]{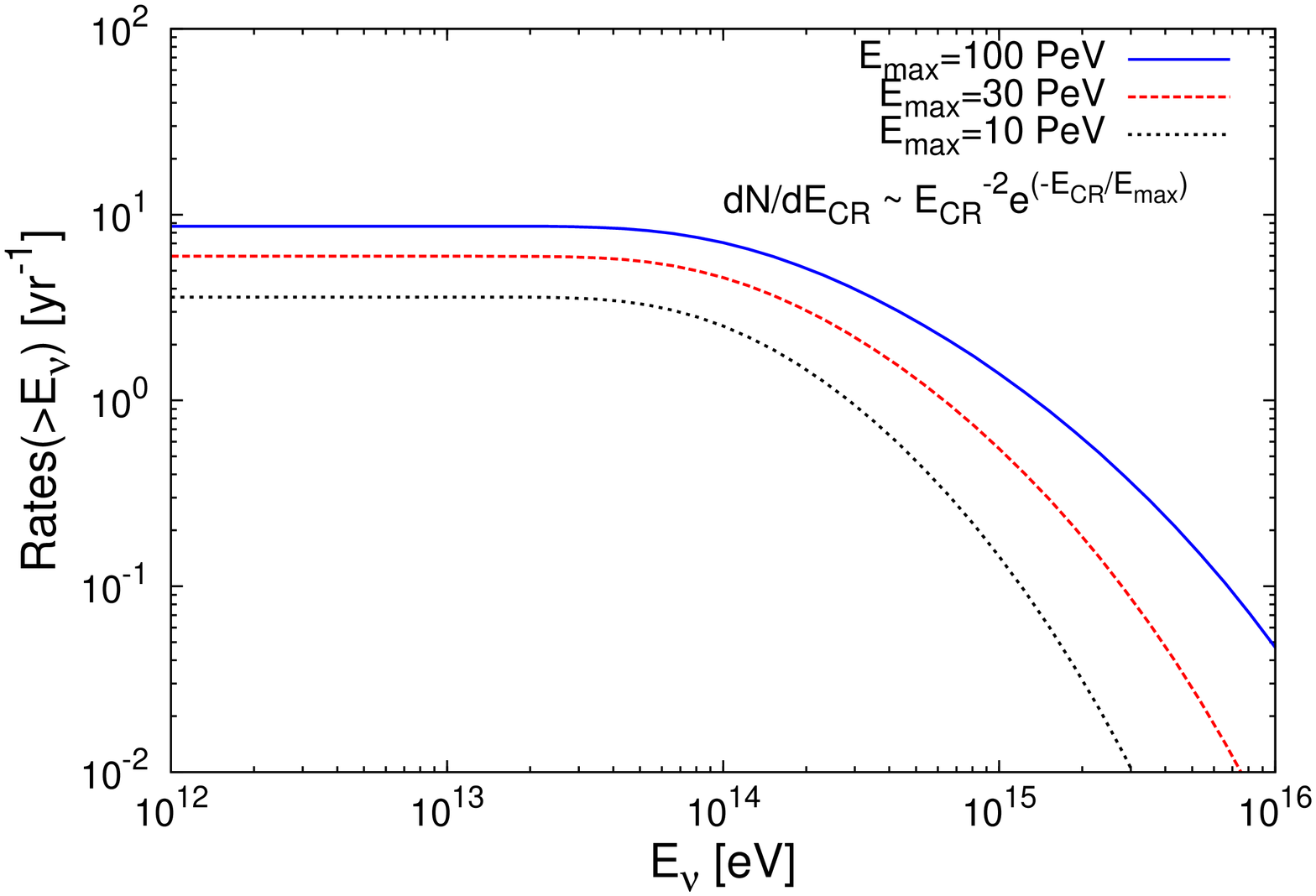}
\caption{TOP: The $\gamma$-ray energy flux for a scenario in which the Fermi bubble
flux is explained by a CR population with shape $dN/dE\propto E^{-2}e^{-E/E_{\rm max}}$, 
with $E_{\rm max}=$10, 30~PeV, and 100~PeV respectively. 
BOTTOM: The corresponding neutrino detection rate expected from the Fermi bubble region
by IceCube.}
\label{Fermi_Bubbles}
\end{figure}

Adopting an energy flux of 100~GeV $\gamma$-rays from the Fermi bubble regions at a 
level of 
$E_{\gamma}^{2}\frac{dN}{dE_{\gamma}}\approx 300{\rm ~eV~cm}^{-2}{\rm ~s}^{-1}{\rm ~sr}^{-1}$,
and accounting for their larger angular size with 
$\Omega_{\rm FB}/\Omega_{\rm d}\approx 8$, 
the results for the Galactic plane region can be scaled up by a factor of $4$
% note- growth in cross-section explains the missing factor of 1.5
to obtain the expected rates from these regions. Thus, as shown in 
fig.~\ref{Fermi_Bubbles}, potentially a rate of 6 events per year would be expected 
to arrive from these regions at energies $>~30$~TeV, in agreement with similar
calculations by others \cite{Adrian-Martinez:2012qpa,Lunardini:2013gva}.

Using the above result, an optimistic estimation for the neutrino luminosity from 
the Fermi bubble regions is,
$L_{\nu}=3\times 10^{36}~{\rm ~erg ~s}^{-1}$. 
However, the long pp cooling time in the region well outside the Galactic 
plane, for which we adopt $n_{p}=10^{-2}{\rm ~cm}^{-3}$, and 
large scale height of 10~kpc, result in an energy transfer efficiency
of CR power into neutrinos,
with $t_{\rm esc}/t_{pp}=(3\times 10^{7})/(4\times 10^{9})=0.008$.
Thus, overall, a proton luminosity with a value of 
$L_{p}\approx 10^{39}{\rm ~erg~s}^{-1}$ is required.
Such a luminosity, though large, is comparable to that required by hadronic origin 
scenarios used to explain the existence of Fermi Bubble regions at GeV energies
\cite{Crocker:2010dg}. Furthermore, should the CR be sufficiently fresh so as not 
yet to have diffusively probed their new environment, the CR spectrum would 
not have steepened through diffusive escape of the higher energy particles.
Shorter residence times within the Fermi bubbles, of course, would increase the
value of the required CR luminosity determined above.

This result, however, follows for the highly optimistic scenario for which the CR 
flux takes an $E^{-2}$ spectral shape over four and a half decades in energy, from 
$\sim $1~TeV up to a cutoff energy of 30~PeV.

{\bf Galactic Halo:} beyond the Fermi bubbles, the diffuse $\gamma$-ray background
\cite{Abdo:2010nz} sits at a level of only a factor of a few lower than the Fermi bubble 
emission flux, and appears isotropic. The origin of this emission remains unclear. 
Furthermore, a dominant component of Fermi bubble emission beyond the observed 
boundaries, with a weaker observed brightness, would be swamped by this background.

With regards a target for pp collisions within the halo, recent new observational evidence 
now suggests that the ``missing baryons problem'' may be solved by the presence of a 
dominant component of baryons in the halo \cite{Gupta:2012rh}. These baryons sit within 
the Galactic virial radius ($\sim 200{\rm ~kpc}$) and may provide an important target 
for Galactic CR in the halo. Assuming $10^{11}~M_{\odot}$ of
baryonic material exists within the halo and is contained within 100~kpc, a mean
density of $\bar{n}_{p}=10^{-3}{\rm ~cm}^{-3}$ is expected.

With the level of the diffuse flux being dictated by the target material distribution,
we adopt a profile of the form 
\begin{eqnarray}
r\frac{dN}{dr}\propto \left(\frac{1.0}{1.0+(r/r_{0})}\right)^{\beta}.
\end{eqnarray}
This expression takes a similar functional form to the MB model in \cite{Fang:2012fk}.

The observed brightness from CR interactions with such a distribution is
dictated by the column depth of material along different lines of sight
convolved with the radial distribution of the CR. Thus, adopting an $r^{-1}$ CR 
distribution for the region $r<r_{0}$, a flat surface brightness would be 
expected for the case of a conical outflow, with the decrease in CR flux being 
compensated by the increase in column depth with distance (ie. $r$) from the Galactic 
center. Beyond $r_{0}$, the observed brightness would be expected to decrease. 
With regards the total emission from shells for this set-up, however, this would
be expected to increase for shells out to $r_{0}$, with emission from larger
shell radii plateauing due to the emitting volume growing with $r^{2}$. It should 
therefore be borne in mind that the Fermi bubbles may be only the tip of an iceberg 
whose true size has yet to be revealed due to the current limits in sensitivity. 
Indeed, recent analysis suggesting an energy dependence of the bubble morphology 
\cite{Yang:2014pia} lends credence to the idea that the present bubble boundaries 
are dictated by instrument sensitivity.

%Particles ejected into the halo by an outflow have escape times actually longer than the interaction time of the CR with the diffuse baryonic matter present in this region.

Assuming that the origin of the full observed neutrino flux comes from a region
with average distance $d_{s}\approx 100{\rm ~kpc}$ away, the observed energy 
flux translates into a source luminosity of 
\begin{eqnarray}
L_{\nu}=4\pi d_{s}^{2}E_{\nu}F_{\nu}=8\times 10^{38}~{\rm ~erg ~s}^{-1}.
\end{eqnarray}

Furthermore, provided that $t_{pp}<t_{\rm esc}$, the target can act as an
energy dump and the observed neutrino flux spectral shape will reflect 
that output by the source. 
Thus, for $K_{\nu}\approx 0.5$, a comparable level efficiency factor 
% CHECK VALUE OF K EXPECTED
is also expected, and the corresponding CR luminosity required to power the 
system is $L_{\rm CR}\approx 10^{39}{\rm ~erg ~s}^{-1}$.
This value is comparable to estimations of the CR source power required 
to support the CR population between the ``knee'' and ``ankle'' regions 
\cite{Gaisser:2005tu}. Indeed, the suggestion of a single source of this 
magnitude powering the CR population above the ``knee'' was made previously 
in \cite{Hillas:1984}.

Alternatively, the flatness in the energy flux of the observed neutrinos
could reflect a weakly or energy independent escape at multi-PeV energies from the 
halo. The associated corresponding decrease in energy transfer efficiency would of 
course require a larger underlying CR luminosity to support the 
observed flux than for the case considered above. However, with recent evidence
indicating a significant budget of underlying power exists for particle acceleration 
within outflows from the Galactic center \cite{Wang:2013dqq}, an increase of more 
than an order of magnitude beyond this estimated luminosity could still be 
considered acceptable.

%DISCUSSION ABOUT ADVECTION + kpc DISTANCE OF INFLUENCE\\

\section{Future Detection}% by KM3Net, HAWC + LHAASO
\label{HAWC_LHAASO}

For the case in which some component of the reported neutrino flux originates
from the Galactic plane region, the future prospects for determining the
validity of such a model are promising. Observations of the Galactic
plane in the near future by the HAWC $\gamma$-ray detector will be able to probe
the multi-TeV brightness of a large fraction of Galactic plane region. 
Such observations will therefore determine whether the Galactic plane flux 
does indeed sit at a level of $\sim 700$~eV~cm$^{-2}$~s$^{-1}$~sr$^{-1}$, as 
motivated by MILAGRO observations of the Cygnus and inner Galactic plane region.
Furthermore, the angular distribution of future IceCube events provides
the most obvious discerning power for such an origin.

For the large-scale diffuse halo scenario, however, the situation is less clear.
At an energy of $\sim$1~PeV, the diffuse CR energy flux sits
at a level of,
%\begin{eqnarray}
$
E_{\rm CR}^{2} dN/dE_{\rm CR}\approx 2\times 10^{5}{\rm ~eV~cm}^{-2}{\rm ~s}^{-1}{\rm sr}^{-1}.
$
%\end{eqnarray}
Assuming that the $\gamma$-ray flux associated with the diffuse neutrino flux is 
diffuse on the largest scales, it is expected to be at 
a level of $E_{\gamma}^{2}dN/dE_{\gamma}\approx 30{\rm ~eV}{\rm ~cm}^{-2}{\rm ~s}^{-1}{\rm sr}^{-1}$,
the photon fraction level of diffuse high energy radiation at PeV energies is
therefore $\gamma/p\sim 10^{-4}$. At lower energies, this ratio decreases even
further due to the rapid growth in the CR energy flux.
The search for the presence of a $\gamma$-ray component in CR air-shower experiments 
via their muon-poor signature presently places a constraint on a diffuse PeV 
$\gamma$-ray flux approximately at this level 
\cite{Matthews:1991,Chantell:1997gs}. Future searches for this component by IceTop 
and IceCube collectively are expected to allow a more sensitive probe of this 
component \cite{Aartsen:2012gka}.

With regards dedicated $\gamma$-ray observatories, in the near future the HAWC detector 
will provide a promising probe for the diffuse scenario, with a capability to detect 
Crab level diffuse fluxes 
($F_{\gamma}(> 1~{\rm TeV}) > 10^{-11} \gamma{\rm ~cm}^{-2}{\rm ~s}^{-1}$) 
from regions less than $\sim$15$^{\circ}$ in size \cite{Abeysekara:2013tza}.
%{\bf CONFIRMATION FROM BRENDA?}. 
Cherenkov telescope experiments such as HESS also have the possibility to probe a diffuse 
background component through their studies of electromagnetic air showers 
\cite{Aharonian:2008aa}. Though unable to discern between electrons and photons, at 
multi-TeV energies, the cooling times of the electrons are extremely short, which 
severely limits their diffusive propagation distance. For this reason, at multi-TeV
energies, electrons from nearby sources are not expected to be detected at Earth, 
and thus the electromagnetic showers seen by HESS are most likely photons.
With regards a detection of diffuse fluxes, at energies beyond $\sim$20~TeV, the
presence of a diffuse electromagnetic background at the level detected by
IceCube should be within reach. Although the fluxes at such energies may not be feasibly
detected with present generation instruments, next generation instruments such as CTA may 
well offer sufficiently sensitivity. In this same vein of next generation instruments,
LHAASO  \cite{Cui:2014bda} also hold great potential for probing diffuse Galactic scenarios
even further. Thus, presently, several promising methods exist for discerning the origin of
the neutrinos, providing complementary additional information for future arrival 
directions studies.

On the other hand, with the halo scenario predicting a potentially very broad angular 
distribution in the arriving neutrino flux, the determination of its origin through
angular distribution studies for this scenario will be challenging.
Furthermore, with PeV $\gamma$-rays being born into the Galactic halo region 
under the above scenario, pair production interactions with the omnipresent 
2.7~K CMB radiation fields is inevitable. Along with the electrons produced 
through charged pion decay, these electrons will preferentially cool via 
synchrotron emission provided the magnetic fields present within the halo are $>\mu$G 
in strength. The energy of this synchrotron emission being 
$E_{\gamma}\approx 50~(E_{e}/{\rm PeV})^{2}(B/{\mu{\rm G}})~{\rm keV}$.
The prospects for detecting this diffuse emission from our own Galaxy are not so
promising, providing only a subdominant component  of the total diffuse X-ray 
background, whose make-up is thought to be dominated by faint extragalactic 
point-like sources \cite{Gilli:2006zi}. 
%Indeed, such emission may provide an important
%component of this background \cite{Protheroe:1980}.
%Similarly, the corresponding diffuse multi-TeV $\gamma$-rays from the 
%subdominant IC cooling channel are expected to be difficult to detect

The possibility of detecting such synchrotron halos around other nearby galaxies, 
the existence of which are motivated by radio observations \cite{Carilli:1992},
are more interesting. Adopting a fiducial distance $3$~Mpc and a luminosity
in PeV electrons of $10^{37}$~erg~s$^{-1}$, the synchrotron energy flux expected from 
such a Galaxy would be
$E_{\gamma}F_{\gamma}\approx 10^{-14}~\left(\frac{L_{e}}{10^{37}{\rm ~erg}{\rm ~s}^{-1}}\right)
\left(\frac{3{\rm ~Mpc}}{D_{s}}\right)^{2}{\rm ~erg}{\rm ~cm}^{-2}{\rm ~s}^{-1}$.
with an angular extension of $\theta\approx 0.1/3\approx 1^{\circ}$.
Thus, for the case in which CR in nearby galactic halos have significantly enhanced
intensities above those present locally in the Milky Way, the detection of synchrotron
halos by new sensitive X-ray instruments such as NuSTAR \cite{Harrison:2013md}
and ASTRO-H \cite{Takahashi:2012jn}
hold great potential. In fact, more powerful galaxies some 30~Mpc away, with larger
expected surface brightness, such as Arp 220, are particularly strongly motivated for 
such studies. Phenomenological predictions of this emission are essentially similar
to those of pair halos expected to exist around AGN, with higher energy electrons
being produced and cooling through synchrotron emission closer to the source region
than lower energy electrons. As a result, a softening of the spectrum is expected
with increasing distance from the source.% REF. TO THIS FOR PAIR HALOS?

%IMPOSSIBLE TO DETECT PHOTONS IN SUCH A BACKGROUND?- THINK NOT

\section{Conclusion}
\label{conclusion}

An investigation of possible Galactic origin scenarios to explain the observation
of multi-TeV to PeV neutrinos reported by IceCube is carried out. 
%Though an origin of these events through $p\gamma$ interactions in compact
%objects like binary systems is also possible, we focus throughout on neutrino 
%production through $pp$ interactions with Galactic material.
On dimensional grounds, the Galactic halo is motivated to be a potentially significant 
source of high energy neutrinos provided that sufficient target material exists out 
at these large radii.

Consideration of constraints from diffuse $\gamma$-ray flux measurements from the
Fermi bubble region by the Fermi satellite, even with extreme extrapolations into
the multi-TeV domain, are demonstrated to yield an insufficient neutrino flux to
account for the excess of neutrinos observed.
An origin of the emission from the more extended Galactic halo region, however, cannot 
be ruled out and may have a physical basis if the neutrino emission is connected to an 
advected CR population. Such a scenario would justify the violation of the uniform CR 
hypothesis usually adopted.

Future detection of either diffuse $\gamma$-rays from the Galactic halo or
synchrotron halos present around our or neighbouring galaxies are suggested as
a means of testing such a Galactic halo hypothesis.

\section*{Acknowledgments}
\label{Acknowledgments}

SG acknowledges support from Agence Nationale de la Recherche under a JCJC grant. 
He also acknowledges the Anton Pannekoek Institute and the GRAPPA Institute at the 
University of Amsterdam for kind hospitality. His work has been partially supported 
by a PHC Van Gogh grant. AT acknowledges a Schroedinger fellowship at DIAS.

%\begin{appendix}
%\section{Effective Area}
%\label{Effective_Area}
%We adopt a monotonic effective area function of the form
%\begin{eqnarray}
%A^{\nu}_{\rm eff}\approx A_{0} \left(\frac{E}{{\rm TeV}}\right)^{\gamma}e^{-(E_{b}/E)}{\rm ~m}^{2}.
%\end{eqnarray}
%with $A_{0}=$1, 0.9, 0.4, $\gamma=$0.4, 0.4, 0.5, and 
%$E_{b}=$117~TeV, 155~TeV, 170~TeV for 
%$\nu_{e}$, $\nu_{\mu}$, and $\nu_{\tau}$, respectively. This parameterisation
%is found to fit well the published all--sky effective area for the three
%different neutrino species shown in fig.~7 of \cite{Aartsen:2013jdh}. 
%A comparison of these descriptions is show in fig.~\ref{Effective_Area_plot}.
%\begin{figure}
%\includegraphics*[width=0.5\linewidth, angle=-90]{Effective_Area.ps}
%\caption{A plot showing the effective area prescription adopted.}
%\label{Effective_Area_plot}
%\end{figure}
%\end{appendix}

\end{document}